\documentclass[12pt,preprint]{aastex}

\begin{document}

\title{Type II Supernovae as Standardized Candles}
\author{Mario Hamuy\altaffilmark{1} \altaffilmark{2}}
\affil{Steward Observatory, The University of Arizona, Tucson, AZ 85721} 
\email{mhamuy@ociw.edu}
\author{Philip A. Pinto}
\affil{Steward Observatory, The University of Arizona, Tucson, AZ 85721}
\email{ppinto@as.arizona.edu}

\altaffiltext{1}{Hubble Fellow}
\altaffiltext{2}{Current Address: The Observatories of the Carnegie Institution of Washington,
813 Santa Barbara Street, Pasadena, CA 91101}

\begin{abstract}
We present evidence for a correlation between expansion velocities of the ejecta of Type II
plateau supernovae and their bolometric luminosities during the plateau phase.
This correlation permits one to standardize the candles and decrease the scatter in the
Hubble diagram from $\sim$1 mag to a level of 0.4 and 0.3 mag in the $V$ and
$I$ bands, respectively. When we restrict the sample to the eight objects which
are well in the Hubble flow ($cz$ $>$ 3,000 km s$^{-1}$) the scatter drops even further
to only 0.2 mag (or 9\% in distance), which is comparable to the
precision yielded by Type Ia supernovae and far better than the
``expanding photosphere method'' applied to Type II supernovae.
Using SN~1987A to calibrate the Hubble diagrams we get $H_0$=55$\pm$12.
\end{abstract}


\section{INTRODUCTION}

\noindent Distances to cosmological objects are the path to get the expansion
rate and the age of the universe. From observations of high-$z$ objects
it is possible to measure how the expansion rate changes with time,
and derive fundamental parameters like the geometry, deceleration, and
energy content of the universe. This experiment has been
recently done by two groups of astronomers using Type Ia supernovae (SNe) 
\citep{riess98,perlmutter99}. Their observations
revealed the surprising result that the universe is presently
accelerating due to a non-zero cosmological constant, a form of dark
energy that permeates space and dominates the total energy content
of the universe, a result largely unanticipated by modern physics.
If confirmed, the result from SNe~Ia would be a revolution in astrophysics.
Before we can fully believe this result we need independent checks.
Although Type II SNe are not as bright as the Ia's, they are the most common
type of supernova and they offer the potential to be used as distance indicators
using models of their atmospheres, a technique known as
the ``expanding photosphere method'' (EPM) \citep{schmidt94,hamuy01}. 
The scatter in the Hubble diagram shows that the precision in an EPM
distance is $\sim$20\% \citep{hamuy01}, which proves significantly larger than the 7\% precision
yielded by SNe~Ia \citep{hamuy96,phillips99}, thus hampering the use of SNe~II for
determination of cosmological parameters.

In principle, the apparent magnitudes of stellar objects can be used to derive distances, 
as long as a class of objects with known luminosities can be identified. Although Type II SNe
display a wide range in luminosities at all epochs -- making it hard to use them as
standard candles -- in this letter we show that the envelope expansion velocities of the
plateau subclass \citep{barbon79} are highly correlated with their luminosities
during the plateau phase. This finding allows us to standardize the candles and
use them to derive distances with precisions comparable to that delivered by SNe~Ia.

\section{OBSERVATIONAL MATERIAL}

Table \ref{tab1} lists the 17 Type II plateau SNe (SNe~II-P) included in this study.
These are all SNe~II-P for which we have 1) precise optical photometry uncontaminated
from host galaxy light, and 2) optical spectroscopy. This table lists the 
specific data sources along with characteristic $VI$ magnitudes and
velocities of the expanding ejecta (derived from the minimum of the Fe II~$\lambda$5169 line
and duly corrected for host galaxy redshifts) for the plateau phase.
Redshifts in the CMB frame come from NED or our own measurements.

\section{LUMINOSITY-VELOCITY RELATION}

\begin{figure}
\figurenum{1}
\epsscale{0.7}
\plotone{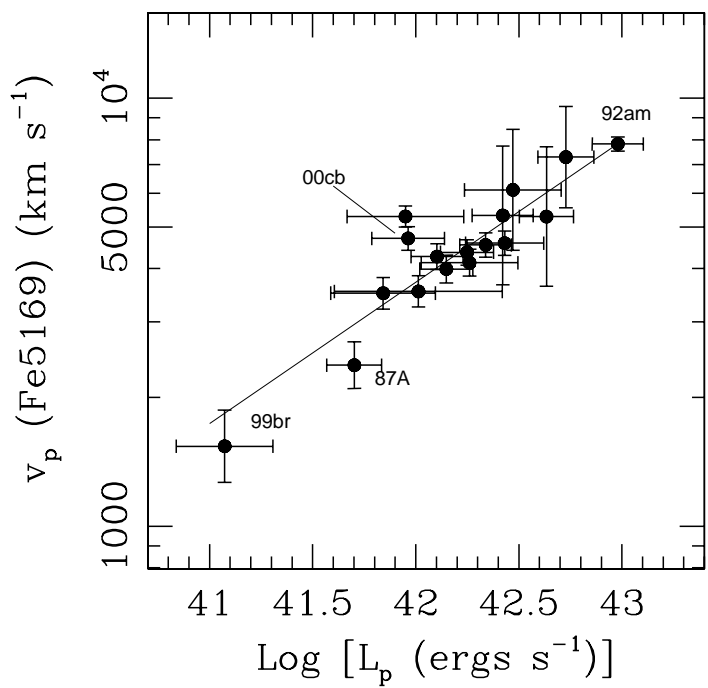}
\caption{Expansion velocities from Fe II $\lambda$5169 versus bolometric luminosity,
both measured in the middle of the plateau (day 50). The ridge line is
a weighted fit to the points and corresponds to $v_p$$\propto$$L_p^{0.33(\pm0.04)}$
(with reduced $\chi^2$ of 0.7).
\label{Lum_vel}}
\end{figure}

Bolometric light curves were derived by H01 for these SNe using 1) $BVI$ photometry, 
2) empirical bolometric corrections, 3) reddening corrections due to our own Galaxy
\citep{schlegel98}, 4) host-galaxy extinction corrections (assuming that all SNe reach
the same color at the end of the plateau), and 5) redshift-based distances ($H_0$=65).
The light curves were all placed in the same time scale using explosion times derived from the
EPM analysis, and considerations about the discovery
and pre-discovery image epochs.
The resulting bolometric luminosities confirmed the well-known fact that SNe~II-P display
a wide range (7.5 mag peak-to-peak) of plateau luminosities. We also noticed that objects
with brighter plateaus have higher envelope expansion velocities.
Figure \ref{Lum_vel} compares the characteristic plateau luminosity
and velocity of these SNe (both measured 50 days after explosion,
which is nearly the middle of the plateau). A remarkable correlation emerges where
SN~1992am and SN~1999br appear as extreme objects with high and low velocities, respectively.
A weighted least-squares fit yields $v_p$$\propto$$L_p^{0.33(\pm0.04)}$, with a reduced $\chi^2$ of 0.7. 

\section{TYPE II SUPERNOVAE AS STANDARD CANDLES}

\begin{figure}
\figurenum{2}
\epsscale{0.7}
\plotone{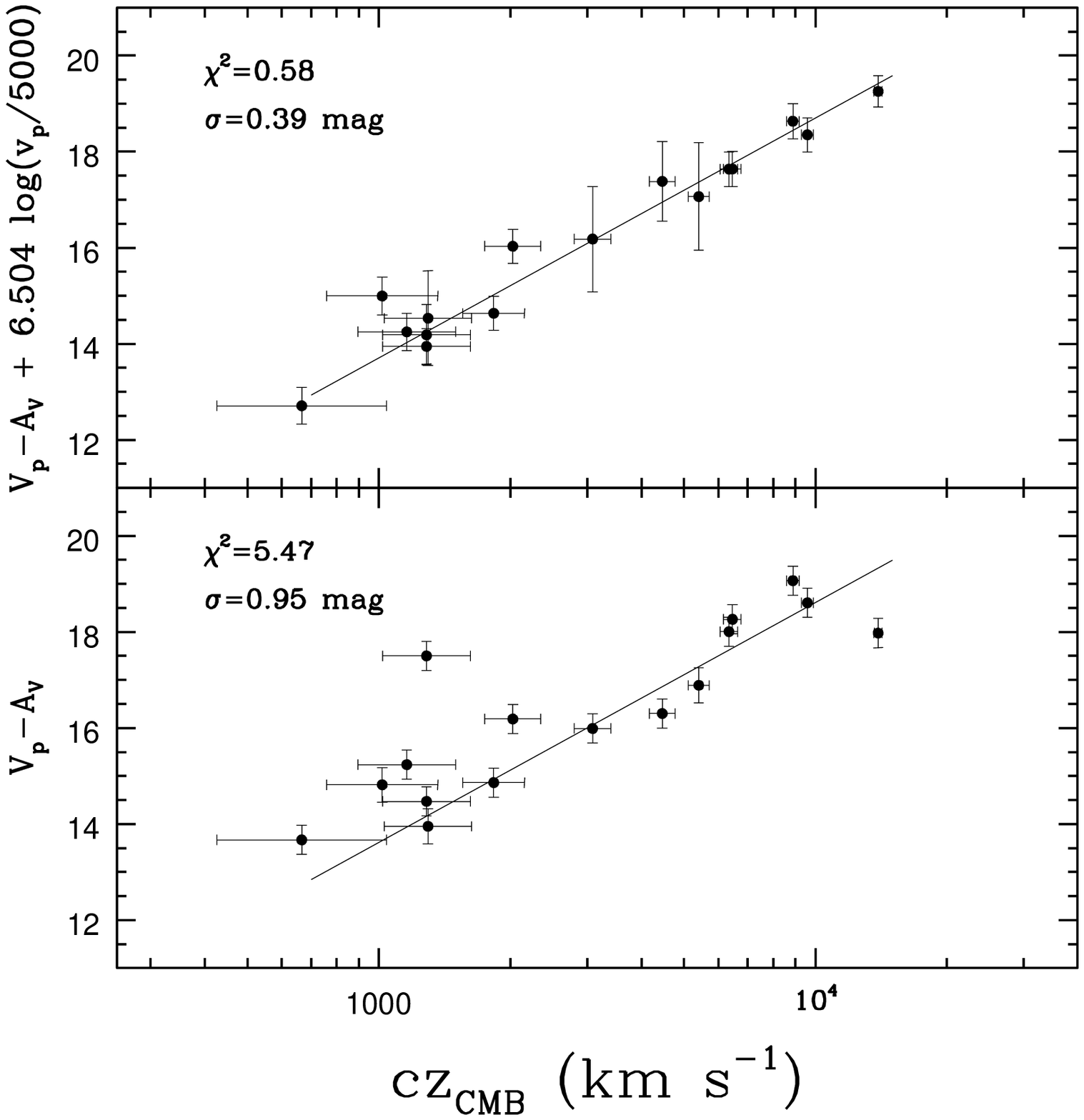}
\caption{(bottom) Raw Hubble diagram from SNe~II plateau $V$ magnitudes. (top) Hubble diagram
from $V$ magnitudes corrected for envelope expansion velocities.
\label{hd3}}
\end{figure}

The tight luminosity-velocity correlation and the inverse square law imply
that the distance to a SN~II-P can be derived from measurements of the apparent
magnitude and envelope velocity.
To test this hypothesis we used $VI$ magnitudes measured on day 50 (corrected for dust
extinction) and expansion velocities derived at the same epoch (see Table \ref{tab1}).
The bottom panel of Figure \ref{hd3} presents the Hubble diagram in the $V$ filter
for all SNe but SN~1987A (which is not in the Hubble flow), while the
top panel shows the same magnitudes after correction for expansion velocities.
A least-squares fit yields the following solution,

\begin{equation}
V_p - A_V + 6.504(\pm0.995)~log (v_p/5000) = 5~log(cz) - 1.294(\pm0.131).
\label{veqn}
\end{equation}

\noindent The scatter drops from 0.95 mag to 0.39 mag,
thus demonstrating that the correction for expansion velocities standardizes
the luminosities of SNe~II significantly. It is interesting to note that most
of the spread  comes from the nearby SNe which are potentially more affected
by peculiar motions of their host galaxies. When we restrict the sample to
the eight objects with $cz$$>$3,000 km s$^{-1}$, the scatter drops to only 0.20 mag.
This implies that the standard candle method can produce relative
distances with a precision of 9\%, which is comparable to the 7\% precision
yielded by SNe~Ia.

\begin{figure}
\figurenum{3}
\epsscale{0.7}
\plotone{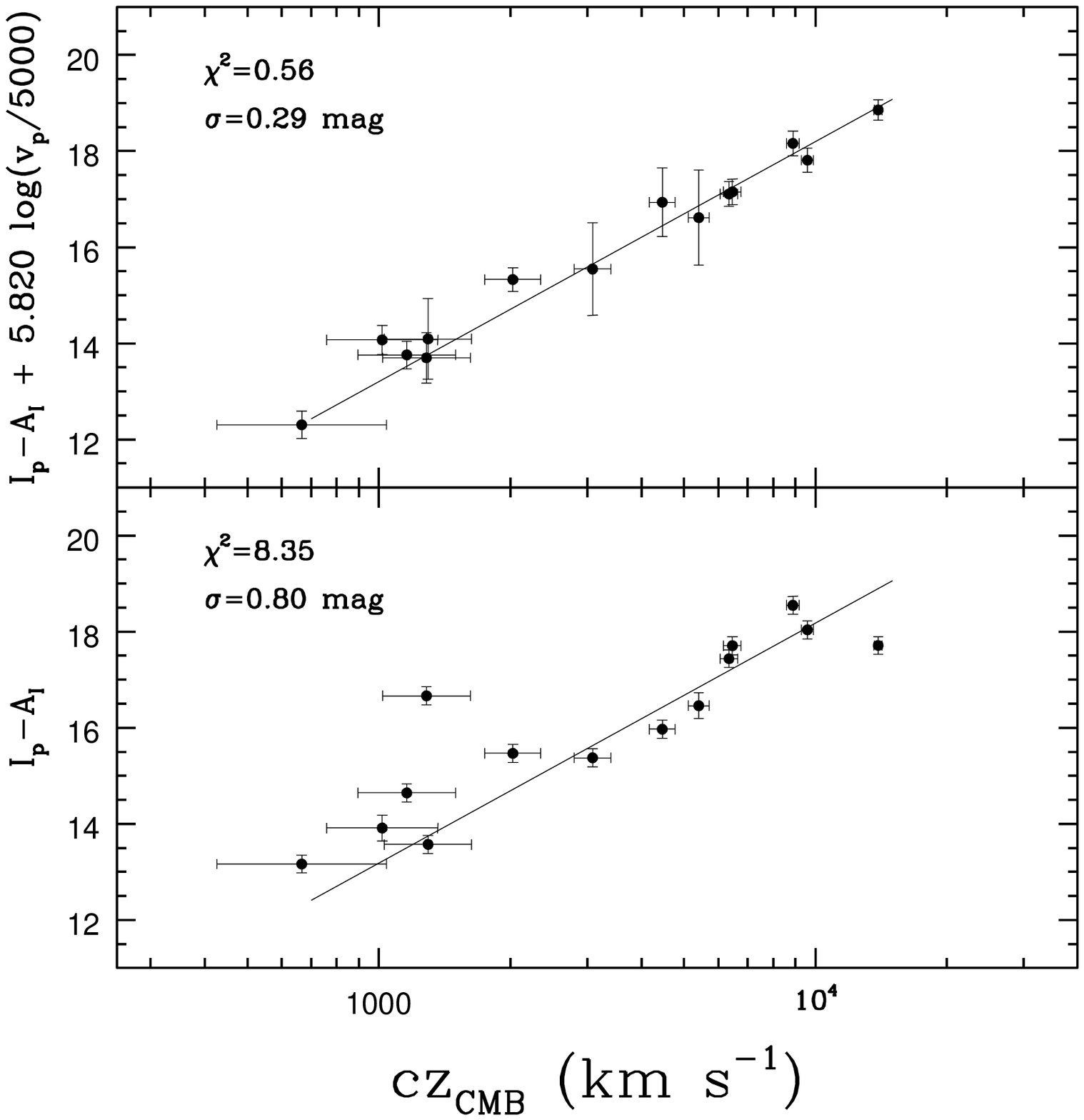}
\caption{(bottom) Raw Hubble diagram from SNe~II plateau $I$ magnitudes. (top) Hubble diagram
from $I$ magnitudes corrected for envelope expansion velocities.
\label{hd4}}
\end{figure}

Figure \ref{hd4} shows the same analysis but in the $I$ band. In this case the
scatter in the raw Hubble diagram is 0.80 mag, which drops to only 0.29 mag
after correction for expansion velocities. This is even smaller that the 0.39
spread in the $V$ band, possibly due to the fact that the effects of dust
extinction are smaller at these wavelengths. The least-squares fit yields the
following solution,

\begin{equation}
I_p - A_I + 5.820(\pm0.764)~log (v_p/5000) = 5~log(cz) - 1.797(\pm0.103).
\label{ieqn}
\end{equation}

\noindent When the eight most distant objects
are employed the spread is 0.21 mag, similar to that obtained from the $V$ magnitudes.

Overall, the standard candle method is characterized by a scatter between 0.39-0.20 mag.
Evidently more objects in the Hubble flow are required to pin down the actual
precision of this technique. The choice of 50 days only has the purpose to represent
approximately the middle of the plateau phase, and it is possible that other
choices could deliver even better results. In its present form 
the method appears very promising for the determination of cosmological distances. Note also that this precision
is better than that yielded by EPM (20\% in distance, or 0.43 mag) \citep{hamuy01} and the
standard candle technique is far less complicated. It only requires a few
spectra and photometry around day 50. A few extra photometric and spectroscopic
observations are required during the plateau in order to solve for dust extinction in the host
galaxy. The time of explosion is also required. Since the duration of the plateau
does not vary much among the different SNe~II, it would suffice to get some photometric
observations during the plateau/nebular phase transition. Alternatively, an EPM
analysis can help at determining $t_0$, but this requires early-time observations.

The standard candle method can be used to solve for the Hubble constant, provided
a distance calibrator is available. Among the objects of our sample,
only SN~1987A has a precise distance in the Cepheid scale.
Assuming an LMC distance of 50 kpc we get $H_0$=54$\pm$13 from
the entire sample of $V$ magnitudes. When we restrict the sample to the eight
most distant objects we get $H_0$=55$\pm$15. The $I$ magnitudes yield $H_0$=53$\pm$10
and $H_0$=56$\pm$12, respectively. These values agree comfortably well with
the 63$\pm$4 value from Cepheids/SNe~Ia \citep{hamuy96,phillips99}. 
Clearly more calibrators are required to improve this estimate, especially
considering that, given its compact progenitor, SN~1987A is not a prototype of the
plateau class and its light curve is quite different than that of typical
plateau events.  It will be interesting to see the results from SN~1999em
which will soon have a Cepheid distance measured with $HST$ (Leonard et al.)

The specific task of checking the cosmic acceleration indicated by SNe~Ia
requires observing SNe~II at the maximum possible redshift. For a typical SN~II
with $M_V$=-17.5 during the plateau phase the apparent magnitude at $z$=0.3 should
be $\sim$23. Discovering such SNe is clearly feasible nowadays. Obtaining such
spectra will be difficult but certainly not impossible with the currently available 8-m class telescopes.
In the worst scenario the standard candle method can produce distance moduli with
a precision of 0.39 mag, so that 13 SNe~II at $z$=0.3 should allow us to measure
the distances of such objects with a precision of 5\% and provide a robust
check on the results of SNe~Ia.

\acknowledgments

\noindent
We thank David Branch for his thorough and constructive referee report.
Support for this work was provided by NASA through Hubble Fellowship grant HST-HF-01139.01-A
awarded by the Space Telescope Science Institute, which is operated by the Association
of Universities for Research in Astronomy, Inc., for NASA, under contract NAS 5-26555.
MH acknowledges financial support from Dave Arnett via Department of Energy grant
DE-FG03-98DP00214/A001.
This research has made use of the NASA/IPAC Extragalactic Database (NED), which is operated by the
Jet Propulsion Laboratory, California Institute of Technology, under
contract with the National Aeronautics and Space Administration.
This research has made use of the SIMBAD database, operated at CDS, Strasbourg, France.


\begin{deluxetable} {lccccccc}
\tablecolumns{8} 
\tablenum{1}
\tablewidth{0pc}
\tablecaption{Redshifts, Magnitudes, and Expansion Velocities of the 17 Type II Supernovae} \label{tab1}
\tablehead{
\colhead{SN} & 
\colhead{$cz(CMB)$} & 
\colhead{$A_{GAL}(V)$} & 
\colhead{$A_{host}(V)$} & 
\colhead{$V_{p}$} & 
\colhead{$I_{p}$} & 
\colhead{$v_{p}$} & 
\colhead{References} \\
\colhead{} &  
\colhead{$\pm$300 km s$^{-1}$} & 
\colhead{} & 
\colhead{$\pm$0.3 mag} & 
\colhead{} & 
\colhead{} & 
\colhead{(km s$^{-1}$)} & 
\colhead{}  }
\startdata
1986L  &  1293 &  0.099  &   0.00 &  14.57(05) &  \nodata    &  4150(300) & 8 \\
1987A  &\nodata&  0.249  &   0.22 &   3.42(05) &  2.45(0.05) &  2391(300) & 4,6 \\
1988A  &  1842 &  0.136  &   0.00 &  15.00(05) &  \nodata    &  4613(300) & 1,7 \\
1990E  &  1023 &  0.082  &   1.00 &  15.90(20) & 14.56(0.20) &  5324(300) & 9,7 \\
1990K  &  1303 &  0.047  &   0.50 &  14.50(20) & 13.90(0.05) &  6142(2000)& 2,7 \\
1991al &  4484 &  0.168  &   0.15 &  16.62(05) & 16.16(0.05) &  7330(2000)& 5,3 \\
1992af &  5438 &  0.171  &   0.00 &  17.06(20) & 16.56(0.20) &  5322(2000)& 5 \\
1992am & 14009 &  0.164  &   0.30 &  18.44(05) & 17.99(0.05) &  7868(300) & 5 \\
1992ba &  1165 &  0.193  &   0.00 &  15.43(05) & 14.76(0.05) &  3523(300) & 5,3 \\
1993A  &  8933 &  0.572  &   0.00 &  19.64(05) & 18.89(0.05) &  4290(300) & 5 \\
1993S  &  9649 &  0.054  &   0.30 &  18.96(05) & 18.25(0.05) &  4569(300) & 5,3 \\
1999br &  1292 &  0.078  &   0.00 &  17.58(05) & 16.71(0.05) &  1545(300) & 5 \\
1999ca &  3105 &  0.361  &   0.30 &  16.65(05) & 15.77(0.05) &  5353(2000)& 5 \\
1999cr &  6376 &  0.324  &   0.00 &  18.33(05) & 17.63(0.05) &  4389(300) & 5 \\
1999eg &  6494 &  0.388  &   0.00 &  18.65(05) & 17.94(0.05) &  4012(300) & 5 \\
1999em &   669 &  0.130  &   0.18 &  13.98(05) & 13.35(0.05) &  3557(300) & 5,10 \\
2000cb &  2038 &  0.373  &   0.00 &  16.56(05) & 15.69(0.05) &  4732(300) & 5 \\
\enddata
\tablerefs{
(1) Benetti et al. 1991;
(2) Capellaro et al. 1995;
(3) Della Valle (unpublished);
(4) Hamuy \& Suntzeff 1990; 
(5) Hamuy 2001;
(6) Phillips et al. 1988;
(7) Phillips (unpublished);
(8) Phillips \& Kirhakos (unpublished);
(9) Schmidt et al. 1993;
(10) Suntzeff et al. (unpublished).}
\end{deluxetable}

\clearpage

\end{document}